\documentclass[preprint,12pt]{elsarticle} 
\usepackage{epsfig}
\usepackage{graphicx}
\usepackage{amssymb,amsfonts,amsmath}

\def\lsim{\mathrel{\rlap{\lower3pt\hbox{\hskip1pt$\sim$}}
     \raise1pt\hbox{$<$}}} 
\def\gsim{\mathrel{\rlap{\lower3pt\hbox{\hskip1pt$\sim$}}
     \raise1pt\hbox{$>$}}} 
\newcommand{\be}{\begin{equation}}
\newcommand{\ee}{\end{equation}}
\newcommand{\bea}{\begin{eqnarray}}
\newcommand{\eea}{\end{eqnarray}}

\begin{document}

\begin{frontmatter}

\title{QCD thermodynamics with dynamical overlap fermions}
\author{
Szabolcs~Bors\'{a}nyi$^a$,
Ydalia Delgado$^b$,
Stephan D\"urr$^{a,c}$,
Zolt\'{a}n~Fodor$^{a,c,d}$, 
S\'{a}ndor~D.~Katz$^d$\footnote{katz@bodri.elte.hu},
Stefan~Krieg$^{a,c}$,
Thomas Lippert$^{a,c}$,
D\'{a}niel N\'ogr\'{a}di$^d$
and
K\'alm\'an~K.~Szab\'o$^a$\\
$^a$Bergische Universit\"at Wuppertal, D-42119 Wuppertal, Germany\\
$^b$Institut f\"ur Physik, Karl-Franzens Universit\"at, Graz, Austria\\
$^c$IAS, J\"ulich Supercomputing Centre, Forschungszentrum J\"ulich, D-52425 J\"ulich, Germany\\
$^d$Institute for Theoretical Physics, E\"otv\"os University, H-1117 Budapest, Hungary
}

\begin{abstract}
We study QCD thermodynamics using two flavors of 
dynamical overlap fermions with quark masses corresponding to a pion mass 
of 350~MeV. We determine several observables on $N_t=6$ and $8$ lattices.
All our runs are performed with fixed global topology. Our results are
compared with staggered ones and a nice agreement is found.
\end{abstract}


\end{frontmatter}

\section{Introduction}

At high temperatures the dominant degrees of freedom of strongly interacting
matter change from hadrons to quarks and gluons. This transition can be studied
using lattice gauge theory. There are various results using different fermion
regularizations. 

Most of these results \cite{Borsanyi:2010bp,Bazavov:2011nk,Borsanyi:2010cj} use the
computationally least expensive staggered discretization which also preserves
some of the continuum chiral symmetry. Even though different staggered results
seem to agree with each other one should not forget that all these works use
the fourth root trick to study $N_f=2+1$ flavors of quarks. There is still an
ongoing debate in the literature about the correctness of this approach.
Furthermore taste symmetry breaking may lead to large discretization errors
when using small quark masses, especially at low temperatures.

There are also several results using Wilson
fermions~\cite{Bornyakov:2009qh,Borsanyi:2011kg,Umeda:2012er}.  Since Wilson
fermions break chiral symmetry explicitly one has to take very fine lattices to
study chiral symmetry restoration at finite temperature. Due to the scattering
of the low lying eigenvalues of the Wilson-Dirac operator one needs large
lattice volumes when going to small pion masses. There are also first results
with twisted mass fermions~\cite{Ilgenfritz:2009ns}.

It seems logical to use chiral fermions to study chiral properties at finite
temperature. Even though lattice chiral fermions are computationally much more
expensive than the other types of discretization there are results in the
literature using domain wall fermions~\cite{Cheng:2009be} as well as first
attempts with overlap fermions~\cite{Cossu:2010rc}. While domain-wall fermions
provide exact chiral symmetry only for an infinite extent of the fifth
dimension, the overlap formulation~\cite{Neuberger:1997fp,Neuberger:1998wv} has
the advantage of exact symmetry on finite four dimensional
lattices~\cite{Luscher:1998pqa}.

In this exploratory Letter we present results using two degenerate flavors
of dynamical overlap fermions. We use two lattice resolutions corresponding
to $N_t=6$ and $8$ temporal extents. We determine the temperature dependence
of the chiral condensate, the
chiral susceptibility, the quark number susceptibility and the Polyakov loop.
The results are compared to $N_f=2$ staggered data, the details of these
simulations are summarized in ~\ref{app_stagg}.

\section{Lattice action and simulation details}
The possibility of using the Hybrid Monte Carlo algorithm (HMC) with overlap
fermions was first discussed in Reference \cite{Fodor:2003bh}. The
overlap operator was implemented with a multi-shift inverter using the
Zolotarev rational approximation~\cite{vandenEshof:2002ms}.
The working of the
algorithm was demonstrated on a small thermodynamic study on $6^3\cdot 4$ 
lattices. It was also observed that treating topology changes requires
special care during the HMC trajectories. One has to track the 
lowest lying eigenvalues of the Wilson kernel of the overlap operator.
This was studied in detail in References \cite{DeGrand:2004nq,Cundy:2005pi,Cundy:2008zc}. 
It was demonstrated in Reference \cite{Egri:2005cx} that one can 
do simulations with a fixed topological charge in several different
sectors and it is possible to determine their relative weight. However,
even this approach requires a tracking of Wilson eigenvalues. 
In Reference \cite{Fukaya:2006vs} 
it was suggested that by adding an extra heavy Wilson fermion to the
action which decouples in the continuum limit, one can disable 
topological sector changes and at the same time speed up the algorithm 
significantly. It was also claimed that in the thermodynamic limit
physics is independent of the global topology and therefore this
approach should give correct results. However, significant 
power-like finite volume corrections  
are expected~\cite{Brower:2003yx,Aoki:2007ka}. 
Here we follow the same approach:
we add an extra Wilson fermion to suppress low lying eigenvalues of
the Wilson kernel and disable tunneling between different topological
sectors. As a further improvement we use smearing in the Wilson kernel.
It was observed in~\cite{Kovacs:2002nz} that smearing significantly improves the 
properties of the overlap operator. Furthermore, since smearing
decreases the eigenvalue density in the middle of the Wilson spectrum
it results in a significant speedup of the algorithm~\cite{Durr:2005an}.

In the gauge sector we use a tree level Symanzik improved gauge
action. The overlap operator can be written as
\be 
D =\left(m_0-\frac{m}{2}\right)\left( 1+\gamma_5 {\rm sgn}\left(H_W\right)\right)+m,
\ee
where $H_W=\gamma_5 D_W$ is the Hermitian Wilson operator with a
negative $-2<-m_0<0$ mass parameter and $m$ is the mass of the overlap quark.
For the Wilson kernel we use two steps of HEX
smearing~\cite{Capitani:2006ni,Durr:2010vn,Durr:2010aw} with smearing parameters of
$\alpha_1=0.72$, $\alpha_2=0.60$ and $\alpha_3=0.44$.  In order to set $m_0$ we
evaluated the Wilson kernel on quenched configurations with the targeted lattice
spacings in this work and located the point which is in the middle between the
physical modes and the first doublers. This resulted in $m_0$=1.3. The simulations
are performed with $N_f=2$ flavors.

As suggested in~\cite{Fukaya:2006vs} we add two irrelevant
terms to the action to suppress low eigenvalues of $H_W$ and fix topology:
\be
S_E=\sum_x\left\{ \bar{\psi}_E(x) D_W(-m_0)\psi_E(x) + \phi^\dagger(x) 
[D_W(-m_0) + i m_B\gamma_5\tau_3]\phi(x) \right\}.
\ee
The first term is the action of two flavors of extra fermions with negative
mass $-m_0$.  The second term, including a two component bosonic field is
included to control the effect of the extra fermions. The eigenvalues of $H_W$
below $m_B$ are most strongly suppressed. Since both $m_0$ and $m_B$ are fixed
in lattice units they correspond to infinitely large masses in the continuum
limit and both these terms decouple.  For the bosonic mass we use $m_B$=0.54.
Since our lattice action results in a fixed topology we aimed to make simulations
with zero topological charge.

We use a HMC algorithm with the Hasenbusch trick~\cite{Hasenbusch:2001ne}, with
an Omelyan integrator~\cite{Takaishi:2005tz} and with a Sexton-Weingarten
multi-scale scheme for the different fields~\cite{Sexton:1992nu}. The latter
ingredient turned out to be rather advantageous, the extra Wilson fermion has
to be integrated with a much smaller stepsize than the much more expensive
overlap fermion.

\begin{figure}
\centerline{\includegraphics*[height=5cm]{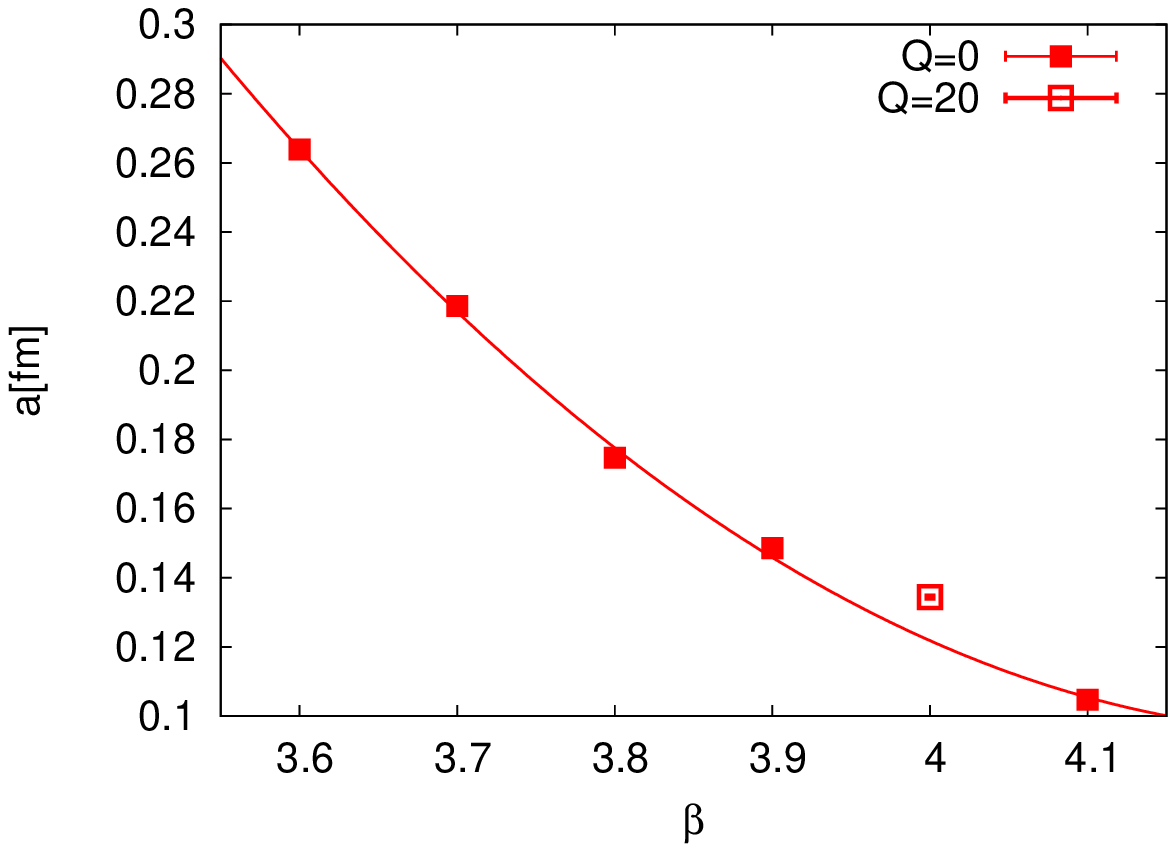}
\includegraphics*[height=5cm]{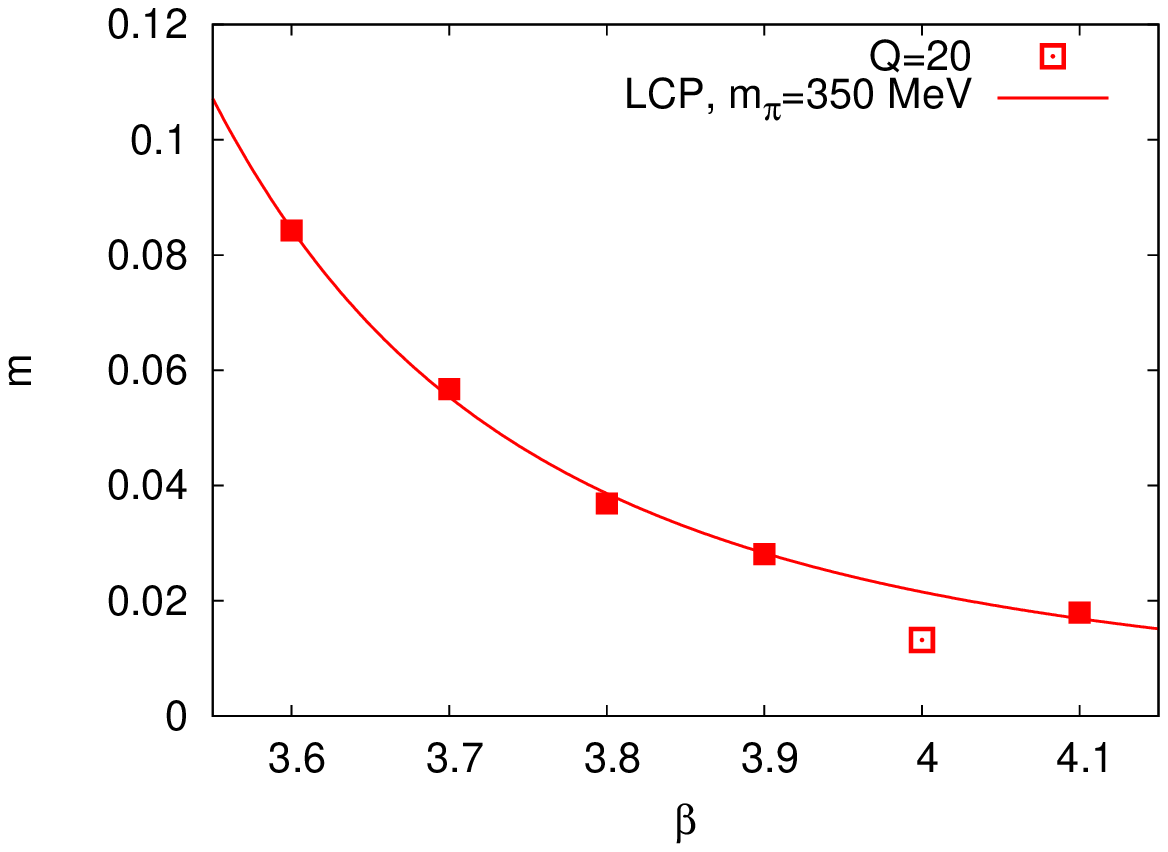}}
\caption{\label{fig_scale} {\em Left:}
The lattice spacing as a function of the $\beta$ coupling.
The opened box shows a run with a fixed 
topological charge of $Q=20$. 
{\em Right:} the bare quark mass as a function of $\beta$.}
\end{figure}

The first step of our analysis was to determine the line of constant physics
(LCP) and the scale. We used $12^3\cdot24$ lattices for $\beta=3.6,3.7,3.8$ and
$3.9$ and $16^3\cdot 32$ for $\beta=4.0$ and $4.1$ with some initial guesses of
the bare quark masses between $0.02$ and $0.06$. We determined the $w_0$
scale~\cite{Borsanyi:2012zs} as well as the pion masses on all of these
lattices. The $w_0$ scale is defined implicitly by the equation 
$\left. d/dt [t^2 E(t)]\right|_{t=w_0^2}=0.3$, where $E(t)$ is the
expectation value of the gauge action evaluated on configurations evolved by
the Wilson flow~\cite{Luscher:2010iy} with parameter $t$. Since due to the chiral symmetry of overlap fermions $m_\pi^2 \propto
m$ and the $w_0$ scale is quite insensitive to the quark mass, it was possible
to tune the quark masses to have a fixed value of $m_\pi\cdot w_0$=0.312 for
each beta without further simulations. With the physical value of $w_0$ at the
$N_f=2+1$ flavor physical point, $w_0=0.1755$~fm, this would correspond to
$m_\pi=350$~MeV. Rigorously a conversion into physical units is only
well defined in QCD with physical quark masses, so the results in MeV or fm are
for orientation only. The lattice spacing as a function of $\beta$ and the LCP are
shown in Figure~\ref{fig_scale}. In 
one of our runs we had a nonzero topological charge, $Q=20$. One can see from the plot
that both the scale and the LCP have still a significant 
dependence on topology for our volumes.

For the finite temperature runs we use two sets of lattices: $12^3\cdot 6$ and
$16^3\cdot 8$. Since the lattice spacing and the LCP is quite ambiguous at
large lattice spacings and the algorithm performs poorly on coarse lattices we
decided not to do $N_t=4$ runs. We had 9 different $\beta$ values for
both $N_t=6$ and $8$ in the range $\beta= 3.6 \dots 4.1$.  For
renormalization we performed runs at the same gauge couplings on $12^4$
lattices up to $\beta=3.7$, on $16^4$ lattices up to $\beta=4.05$ and 
on $24^3\cdot 32$ at $\beta=4.1$. We collected up to 2000 HMC trajectories 
for each finite temperature, and around 500 for each zero temperature point.

\section{Results}
\begin{figure}
\centerline{\includegraphics*[height=7cm]{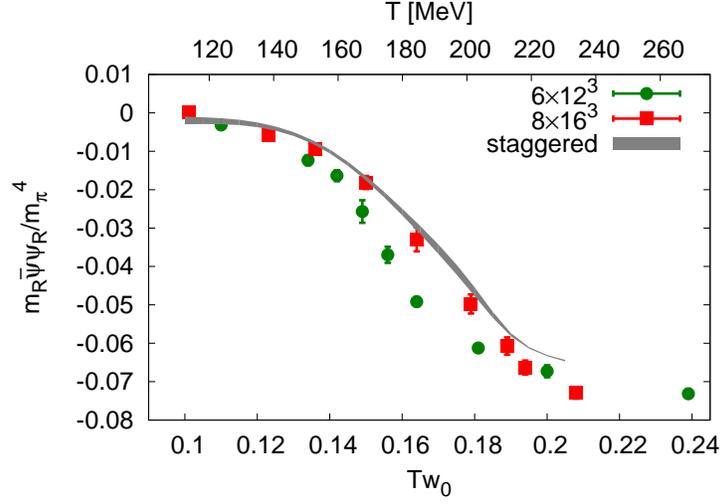}}
\caption{\label{fig_pbp}
The renormalized chiral condensate as a function of temperature on $N_t=6$ and $8$ lattices. 
The upper temperature scale is for illustration only and it is based
$w_0=0.1755$~fm~\cite{Borsanyi:2012zs}. The gray band shows our staggered
estimate based on $N_t=6,8$ and 10 simulations. For details see ~\ref{app_stagg}.
}
\end{figure}

The first quantity we study is the {\em chiral condensate}, 
$\bar{\psi}\psi=(T/V) \partial/\partial m \log Z$. This can be renormalized
using the zero temperature condensate $\bar{\psi}\psi_0$
(this observable was studied in \cite{Endrodi:2011gv}):
\be
m_R\bar{\psi}\psi_R/m_\pi^4=m(\bar{\psi}\psi -\bar{\psi}\psi_0)/m_{\pi}^4
\ee
The renormalized condensate 
is plotted in Figure~\ref{fig_pbp} together with our staggered estimate. 
The temperatures are converted to MeV
again by using the physical value of $w_0$. We can see that there is still
some lattice spacing dependence, but the $N_t=8$ results are very close
to the staggered ones. One observes a broad cross-over, 
similar to the staggered results at physical quark masses~\cite{Aoki:2006we}.

\begin{figure}
\centerline{\includegraphics*[height=7cm]{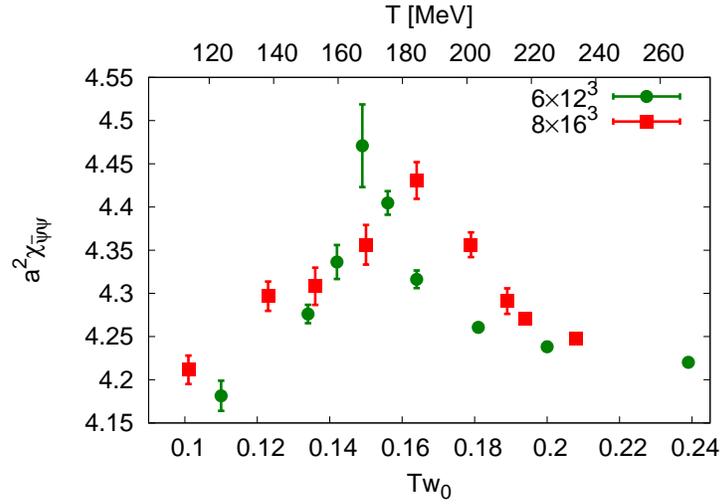}}
\caption{\label{fig_susc}
The bare chiral susceptibility as a function of temperature on our
$N_t=6$ and $8$ lattices. 
}
\end{figure}

We have also determined the {\em chiral susceptibility}
\be
\chi_{\bar{\psi}\psi}=(T/V)\partial^2/ \partial m^2 \log Z,
\ee
but at the present level of our statistics renormalization resulted
in large errors. Therefore we only show the bare susceptibilities
in Figure~\ref{fig_susc}. Even though the results obtained at our 
two lattice spacings cannot be compared
directly, the peaks nicely signal the transition and we can see that the
transition temperature defined from this quantity has small lattice spacing
dependence.

\begin{figure}
\centerline{\includegraphics*[height=7cm]{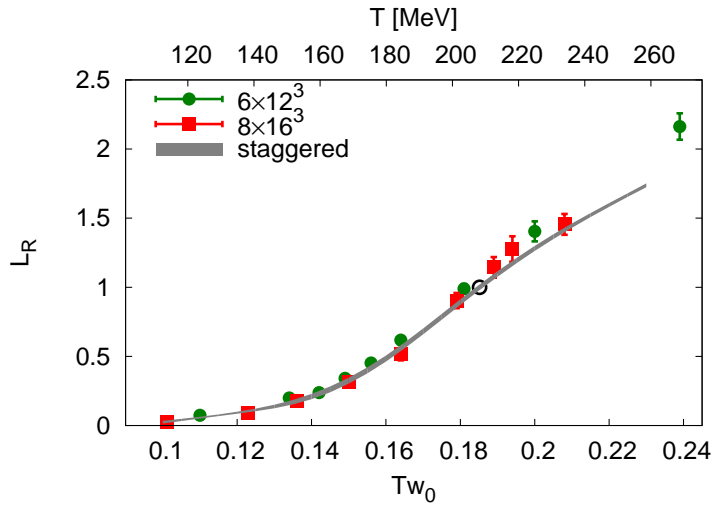}}
\caption{\label{fig_L}
The renormalized Polyakov loop as a function of temperature on $N_t=6$ and $8$ 
lattices as well as the staggered result. The black circle 
represents our renormalization condition, $L_R(T=208 {\rm MeV})=1$.
}
\end{figure}

The next quantity we study is the {\em Polyakov loop}. The bare quantity has a
multiplicative divergence of the form $\exp[F_0(\beta)/T]$ where the 
divergent term $F_0$ can be determined up to a constant~\cite{Aoki:2006br}. 
Different constants
correspond to different renormalization schemes. We determine $F_0$ entirely
from finite temperature simulations in the following way. 
We perform runs with six different $N_t$ values, $N_t=4,5,6,7,8$ and $9$, all
lattice extents are $16^3\cdot N_t$.
We choose a fixed physical temperature such that these $N_t$ values span
our $\beta$ range. This corresponds to a temperature of $208$~MeV 
(having $N_t=9$ at $\beta=4.1$). From these runs we can determine
$F_0(\beta)=1/N_t\cdot\log L$ at six $\beta$ values. This can then be
extended by interpolation to all of our couplings. This renormalization
scheme corresponds to the condition $L_R(T=208{\rm MeV})=1$.
The renormalized
Polyakov loop is then given as:
\be
L_R=L_0 e^{-N_t\cdot F_0(\beta)},
\ee
where $L_0$ is the bare Polyakov loop.
The result is shown in Figure~\ref{fig_L}. We can see
almost no lattice spacing dependence and an excellent agreement with the 
staggered results. 

\begin{figure}
\centerline{\includegraphics*[height=7cm]{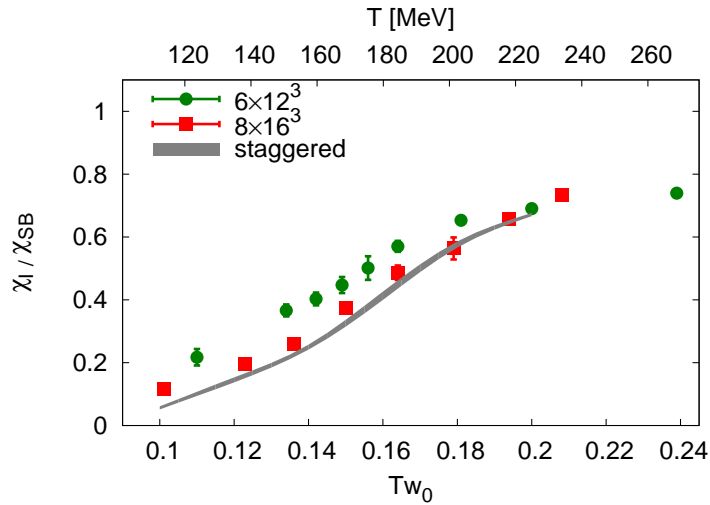}}
\caption{\label{fig_qns}
The isospin susceptibility as a function of temperature on $N_t=6$ and $8$ lattices. The
gray band again shows the corresponding staggered result.
All data sets were normalized by their corresponding SB limits (cf. Table \ref{tab_SB}).
}
\end{figure}

Our final observable is the {\em isospin susceptibility}, 
\be\label{eq_qsusc}
\chi_{I}=\left. (T/V) \partial^2/ \partial \mu_I^2 \log Z\right|_{\mu_I=0},
\ee 
where $\mu_I$ is the isospin
chemical potential, i.e. the quark chemical potentials are $\mu_u=\mu_I/2$ and
$\mu_d=-\mu_I/2$. Obtaining results at non-vanishing chemical 
potentials is very CPU
demanding 
(see e.g.~\cite{Fodor:2001au,Fodor:2004nz}). Even though
our analysis to $\mu>0$ is beyond the scope of the
present Letter, including the chemical potential even on the level of 
eq. (\ref{eq_qsusc}) is quite interesting. The reason is that there is an 
ongoing discussion in the literature about the proper inclusion
of the chemical potential in the overlap operator~\cite{Bloch:2006cd,Gavai:2009vb,Narayanan:2011ff}. We follow Reference \cite{Bloch:2006cd}
and define the chemical potential as a fourth, imaginary component of the 
temporal gauge field and use the generalization of the sign function:
${\rm sgn}(z)={\rm sgn Re}(z)$. The second derivative can be calculated
using the formulas of ~\ref{app_zol}.
As a tree level improvement we normalize all susceptibilities
with the corresponding Stefan-Boltzmann (SB) values which, for our choice
of $m_0=1.3$ are given in Table~\ref{tab_SB} for a number of $N_t$'s
both for infinite volume and our relatively small aspect ratio, $\xi=2$.
\begin{table}
\begin{center}
\begin{tabular}{|c|c|c|c|c|c|}
\hline
$N_t$&4&6&8&10&12\\
\hline\hline
$\xi=2$ overlap     & 1.700 & 1.588 & 1.362 & 1.241 & 1.186 \\
\hline
$\xi=\infty$ overlap& 1.619 & 1.513 & 1.290  & 1.170 & 1.117  \\
\hline
$\xi=\infty$ staggered    & 2.235 & 1.861 & 1.473 & 1.266 & 1.164 \\
\hline
$\xi=\infty$ Wilson& 4.168 & 2.258  & 1.521 & 1.265 &  1.161 \\
\hline
\end{tabular}\end{center}
\caption{\label{tab_SB} 
Stefan-Boltzmann limits of the quark number susceptibility for three colors of
overlap quarks with $m_0=1.3$ for aspect ratios ($\xi$) of $2$ and infinity. As a 
comparison we also give the infinite volume values for Wilson and staggered quarks.
}
\end{table}
Our results for the quark number susceptibility are shown in Figure~\ref{fig_qns}.
Again, the $N_t=8$ results are very close to the staggered ones.

\section{Conclusions, outlook}
We presented results for the temperature dependence of several observables
using dynamical overlap fermions. Our results show that on $N_t=6$ and
$8$ lattices cutoff effects are still present but not severe. The comparison
with staggered results obtained on $N_t=6$,8 and 10 lattices is quite
encouraging: there is a good chance for a reliable continuum extrapolation
from $N_t\lsim 10$ lattices with overlap fermions.

There are several remaining issues to investigate in future studies. A particularly
interesting question is the dependence of the results on the global topology
and how it disappears in the thermodynamic limit. Such analysis requires 
a series of runs on larger volumes. To approach the continuum limit one needs
at least one more lattice spacing. Finally, one might also include the
strange quark and decrease the light quark masses to their physical values.

\section*{Acknowledgments}
Computations were carried out on 
GPU~\cite{Egri:2006zm} clusters at the
Universities of Wuppertal and Budapest as well as on supercomputers in
For\-schungs\-zentrum Juelich.  This work is supported in part by the Deutsche
Forschungsgemeinschaft grants FO 502/2 and SFB- TR 55 and by the EU
(FP7/2007-2013)/ERC No. 208740.

\appendix
\section{Derivatives of the Zolotarev approximation}\label{app_zol}

Here we present results on the first and second derivatives of the sign
function. In the following the $\delta$ symbol stands for a general derivation
operator, for example a derivation with respect to the gauge links or with
respect to the chemical potential. The Zolotarev approximation of the sign
function of an operator $h$ can be written in a partial fraction expansion as:
\be
{\rm sgn}(h)\approx h\left( c_0 + \sum_i \frac{c_i}{h^2+q_i} \right)= h \left( c_0 + \sum_i c_i Q_i \right),
\ee
where $Q_i=(h^2+q_i)^{-1}$ was introduced.
The first derivative is
\be
\delta {\rm sgn}(h)\approx \delta h \left( c_0 + \sum_i c_i Q_i \right) - \sum_i c_i Q_i h \delta h^2 Q_i,
\ee
whereas the second derivative is
\be
\delta^2 {\rm sgn}(h) \approx
\delta^2 h \left( c_0 + \sum_i c_i Q_i \right)+
\sum_i Q_i \left[ 2\delta h^2 Q_i h \delta h^2 - h\delta^2 h^2 -2\delta h \delta h^2\right] c_i Q_i.
\ee

\section{Details of the staggered calculations} \label{app_stagg}
In this appendix we describe our staggered analysis. The sole goal
of this study was to provide a basis of comparison for the overlap data.
To suppress most of the taste-breaking effects of the staggered action
we used four levels of stout smearings ($\rho=0.125$) in the fermionic
sector, which contained here the two light quarks only.
We determined a two-flavor LCP
for this staggered action using the same definition as in the case of the
overlap simulations.

To this end we simulated sixteen ensembles (four lattice spacings with four
quark masses each). The smallest mass was always close to the final LCP's value.
In the range $\beta=3.8--4.1$ we could fit the emerging $am_q(\beta)$
and $w_0/a(\beta)$ functions to sub-percent accuracy.  

For the renormalization of the chiral condensate we made a chiral interpolation
(or extrapolation for some lattice spacings) of the vacuum condensate
$\langle\bar\psi\psi\rangle_0$. 

The Polyakov loop requires renormalization, too, here we calculated the
static potential for our sixteen $T=0$ ensembles.
One can select any physical distance to remove all divergences. Since
$w_0$ is our most accurately known scale, we used $\sqrt{8}w_0$
which is around 0.5~fm and 
defined $Z=\exp(V(r=\sqrt{8}w_0)/2)$. 
After this renormalization our final estimate for
the Polyakov loop was multiplied by an $\exp(-\delta/(w_0T))$ function, which
facilitates finite transformations between different renormalization schemes.
We set $\delta=0.1566$ to approximately reproduce the
renormalization condition used in the overlap results.  The renormalization
procedure might seem elaborate, yet the dominant source of
error comes from our finite temperature statistics.

Our finite temperature simulations were performed with the same aspect
ratio as with the overlap action, but here, in addition to $N_t=6$ and 8,
we also made a set of $N_t=10$ ensembles. In most of the temperature range
we had two lattice spacings. The agreement between the results at these
lattice spacings was even slightly better than in our previous work in full
QCD~\cite{Aoki:2006br}. For our ``staggered estimate'' 
we use the $N_t=10$ at low temperatures and we present $N_t=8$ data 
starting from a temperature
where both $N_t=8$ and 10 data were present and they were in agreement.

\bibliographystyle{model1-num-names}
\bibliography{ov_thermo}

\begin{thebibliography}{40}
\expandafter\ifx\csname natexlab\endcsname\relax\def\natexlab#1{#1}\fi
\providecommand{\bibinfo}[2]{#2}
\ifx\xfnm\relax \def\xfnm[#1]{\unskip,\space#1}\fi
\bibitem[{Borsanyi et~al.(2010)}]{Borsanyi:2010bp}
\bibinfo{author}{S.~Borsanyi}, et~al.,
\newblock \bibinfo{title}{{Is there still any Tc mystery in lattice QCD?
  Results with physical masses in the continuum limit III}},
\newblock \bibinfo{journal}{JHEP} \bibinfo{volume}{09} (\bibinfo{year}{2010})
  \bibinfo{pages}{073}.
\bibitem[{Bazavov et~al.(2012)}]{Bazavov:2011nk}
\bibinfo{author}{A.~Bazavov}, et~al.,
\newblock \bibinfo{title}{{The chiral and deconfinement aspects of the QCD
  transition}},
\newblock \bibinfo{journal}{Phys. Rev.} \bibinfo{volume}{D85}
  (\bibinfo{year}{2012}) \bibinfo{pages}{054503}.
\bibitem[{Borsanyi et~al.(2010)}]{Borsanyi:2010cj}
\bibinfo{author}{S.~Borsanyi}, et~al.,
\newblock \bibinfo{title}{{The QCD equation of state with dynamical quarks}},
\newblock \bibinfo{journal}{JHEP} \bibinfo{volume}{11} (\bibinfo{year}{2010})
  \bibinfo{pages}{077}.
\bibitem[{Bornyakov et~al.(2010)}]{Bornyakov:2009qh}
\bibinfo{author}{V.~G. Bornyakov}, et~al.,
\newblock \bibinfo{title}{{Probing the finite temperature phase transition with
  Nf=2 nonperturbatively improved Wilson fermions}},
\newblock \bibinfo{journal}{Phys. Rev.} \bibinfo{volume}{D82}
  (\bibinfo{year}{2010}) \bibinfo{pages}{014504}.
\bibitem[{Borsanyi et~al.(2011)}]{Borsanyi:2011kg}
\bibinfo{author}{S.~Borsanyi}, et~al.,
\newblock \bibinfo{title}{{QCD thermodynamics with Wilson fermions}}
  (\bibinfo{year}{2011}).
\bibitem[{Umeda et~al.(2012)}]{Umeda:2012er}
\bibinfo{author}{T.~Umeda}, et~al.,
\newblock \bibinfo{title}{{Equation of state in 2+1 flavor QCD with improved
  Wilson quarks by the fixed scale approach}}  (\bibinfo{year}{2012}).
\bibitem[{Ilgenfritz et~al.(2009)}]{Ilgenfritz:2009ns}
\bibinfo{author}{E.~M. Ilgenfritz}, et~al.,
\newblock \bibinfo{title}{{Phase structure of thermal lattice QCD with $N_f=2$
  twisted mass Wilson fermions}},
\newblock \bibinfo{journal}{Phys. Rev.} \bibinfo{volume}{D80}
  (\bibinfo{year}{2009}) \bibinfo{pages}{094502}.
\bibitem[{Cheng et~al.(2010)}]{Cheng:2009be}
\bibinfo{author}{M.~Cheng}, et~al.,
\newblock \bibinfo{title}{{The finite temperature QCD using 2+1 flavors of
  domain wall fermions at $N_t = 8$}},
\newblock \bibinfo{journal}{Phys. Rev.} \bibinfo{volume}{D81}
  (\bibinfo{year}{2010}) \bibinfo{pages}{054510}.
\bibitem[{Cossu et~al.(2010)}]{Cossu:2010rc}
\bibinfo{author}{G.~Cossu}, et~al.,
\newblock \bibinfo{title}{{Finite temperature QCD at fixed Q with overlap
  fermions}},
\newblock \bibinfo{journal}{PoS} \bibinfo{volume}{LATTICE2010}
  (\bibinfo{year}{2010}) \bibinfo{pages}{174}.
\bibitem[{Neuberger(1998{\natexlab{a}})}]{Neuberger:1997fp}
\bibinfo{author}{H.~Neuberger},
\newblock \bibinfo{title}{{Exactly massless quarks on the lattice}},
\newblock \bibinfo{journal}{Phys. Lett.} \bibinfo{volume}{B417}
  (\bibinfo{year}{1998}{\natexlab{a}}) \bibinfo{pages}{141--144}.
\bibitem[{Neuberger(1998{\natexlab{b}})}]{Neuberger:1998wv}
\bibinfo{author}{H.~Neuberger},
\newblock \bibinfo{title}{{More about exactly massless quarks on the lattice}},
\newblock \bibinfo{journal}{Phys. Lett.} \bibinfo{volume}{B427}
  (\bibinfo{year}{1998}{\natexlab{b}}) \bibinfo{pages}{353--355}.
\bibitem[{Luscher(1998)}]{Luscher:1998pqa}
\bibinfo{author}{M.~Luscher},
\newblock \bibinfo{title}{{Exact chiral symmetry on the lattice and the
  Ginsparg- Wilson relation}},
\newblock \bibinfo{journal}{Phys. Lett.} \bibinfo{volume}{B428}
  (\bibinfo{year}{1998}) \bibinfo{pages}{342--345}.
\bibitem[{Fodor et~al.(2004)Fodor, Katz, and Szabo}]{Fodor:2003bh}
\bibinfo{author}{Z.~Fodor}, \bibinfo{author}{S.~D. Katz},
  \bibinfo{author}{K.~K. Szabo},
\newblock \bibinfo{title}{{Dynamical overlap fermions, results with hybrid
  Monte- Carlo algorithm}},
\newblock \bibinfo{journal}{JHEP} \bibinfo{volume}{08} (\bibinfo{year}{2004})
  \bibinfo{pages}{003}.
\bibitem[{van~den Eshof et~al.(2002)van~den Eshof, Frommer, Lippert, Schilling,
  and van~der Vorst}]{vandenEshof:2002ms}
\bibinfo{author}{J.~van~den Eshof}, \bibinfo{author}{A.~Frommer},
  \bibinfo{author}{T.~Lippert}, \bibinfo{author}{K.~Schilling},
  \bibinfo{author}{H.~van~der Vorst},
\newblock \bibinfo{title}{{Numerical methods for the QCD overlap operator. I.
  Sign function and error bounds}},
\newblock \bibinfo{journal}{Comput.Phys.Commun.} \bibinfo{volume}{146}
  (\bibinfo{year}{2002}) \bibinfo{pages}{203--224}.
\bibitem[{DeGrand and Schaefer(2005)}]{DeGrand:2004nq}
\bibinfo{author}{T.~A. DeGrand}, \bibinfo{author}{S.~Schaefer},
\newblock \bibinfo{title}{{Physics issues in simulations with dynamical overlap
  fermions}},
\newblock \bibinfo{journal}{Phys. Rev.} \bibinfo{volume}{D71}
  (\bibinfo{year}{2005}) \bibinfo{pages}{034507}.
\bibitem[{Cundy et~al.(2009{\natexlab{a}})}]{Cundy:2005pi}
\bibinfo{author}{N.~Cundy}, et~al.,
\newblock \bibinfo{title}{{Numerical methods for the QCD overlap operator. IV:
  Hybrid Monte Carlo}},
\newblock \bibinfo{journal}{Comput. Phys. Commun.} \bibinfo{volume}{180}
  (\bibinfo{year}{2009}{\natexlab{a}}) \bibinfo{pages}{26--54}.
\bibitem[{Cundy et~al.(2009{\natexlab{b}})Cundy, Krieg, Lippert, and
  Schafer}]{Cundy:2008zc}
\bibinfo{author}{N.~Cundy}, \bibinfo{author}{S.~Krieg},
  \bibinfo{author}{T.~Lippert}, \bibinfo{author}{A.~Schafer},
\newblock \bibinfo{title}{{Topological tunneling with Dynamical overlap
  fermions}},
\newblock \bibinfo{journal}{Comput. Phys. Commun.} \bibinfo{volume}{180}
  (\bibinfo{year}{2009}{\natexlab{b}}) \bibinfo{pages}{201--208}.
\bibitem[{Egri et~al.(2006)Egri, Fodor, Katz, and Szabo}]{Egri:2005cx}
\bibinfo{author}{G.~I. Egri}, \bibinfo{author}{Z.~Fodor},
  \bibinfo{author}{S.~D. Katz}, \bibinfo{author}{K.~K. Szabo},
\newblock \bibinfo{title}{{Topology with dynamical overlap fermions}},
\newblock \bibinfo{journal}{JHEP} \bibinfo{volume}{01} (\bibinfo{year}{2006})
  \bibinfo{pages}{049}.
\bibitem[{Fukaya et~al.(2006)}]{Fukaya:2006vs}
\bibinfo{author}{H.~Fukaya}, et~al.,
\newblock \bibinfo{title}{{Lattice gauge action suppressing near-zero modes of
  H(W)}},
\newblock \bibinfo{journal}{Phys. Rev.} \bibinfo{volume}{D74}
  (\bibinfo{year}{2006}) \bibinfo{pages}{094505}.
\bibitem[{Brower et~al.(2003)Brower, Chandrasekharan, Negele, and
  Wiese}]{Brower:2003yx}
\bibinfo{author}{R.~Brower}, \bibinfo{author}{S.~Chandrasekharan},
  \bibinfo{author}{J.~W. Negele}, \bibinfo{author}{U.~Wiese},
\newblock \bibinfo{title}{{QCD at fixed topology}},
\newblock \bibinfo{journal}{Phys.Lett.} \bibinfo{volume}{B560}
  (\bibinfo{year}{2003}) \bibinfo{pages}{64--74}.
\bibitem[{Aoki et~al.(2007)Aoki, Fukaya, Hashimoto, and Onogi}]{Aoki:2007ka}
\bibinfo{author}{S.~Aoki}, \bibinfo{author}{H.~Fukaya},
  \bibinfo{author}{S.~Hashimoto}, \bibinfo{author}{T.~Onogi},
\newblock \bibinfo{title}{{Finite volume QCD at fixed topological charge}},
\newblock \bibinfo{journal}{Phys.Rev.} \bibinfo{volume}{D76}
  (\bibinfo{year}{2007}) \bibinfo{pages}{054508}.
\bibitem[{Kovacs(2003)}]{Kovacs:2002nz}
\bibinfo{author}{T.~G. Kovacs},
\newblock \bibinfo{title}{{Locality and topology with fat link overlap
  actions}},
\newblock \bibinfo{journal}{Phys. Rev.} \bibinfo{volume}{D67}
  (\bibinfo{year}{2003}) \bibinfo{pages}{094501}.
\bibitem[{Durr et~al.(2005)Durr, Hoelbling, and Wenger}]{Durr:2005an}
\bibinfo{author}{S.~Durr}, \bibinfo{author}{C.~Hoelbling},
  \bibinfo{author}{U.~Wenger},
\newblock \bibinfo{title}{{Filtered overlap: Speedup, locality, kernel
  non-normality and Z(A) =~ 1}},
\newblock \bibinfo{journal}{JHEP} \bibinfo{volume}{0509} (\bibinfo{year}{2005})
  \bibinfo{pages}{030}.
\bibitem[{Capitani et~al.(2006)Capitani, Durr, and Hoelbling}]{Capitani:2006ni}
\bibinfo{author}{S.~Capitani}, \bibinfo{author}{S.~Durr},
  \bibinfo{author}{C.~Hoelbling},
\newblock \bibinfo{title}{{Rationale for UV-filtered clover fermions}},
\newblock \bibinfo{journal}{JHEP} \bibinfo{volume}{0611} (\bibinfo{year}{2006})
  \bibinfo{pages}{028}. \bibinfo{note}{26 pages, 5 figures}.
\bibitem[{Durr et~al.(2011{\natexlab{a}})}]{Durr:2010vn}
\bibinfo{author}{S.~Durr}, et~al.,
\newblock \bibinfo{title}{{Lattice QCD at the physical point: light quark
  masses}},
\newblock \bibinfo{journal}{Phys. Lett.} \bibinfo{volume}{B701}
  (\bibinfo{year}{2011}{\natexlab{a}}) \bibinfo{pages}{265--268}.
\bibitem[{Durr et~al.(2011{\natexlab{b}})}]{Durr:2010aw}
\bibinfo{author}{S.~Durr}, et~al.,
\newblock \bibinfo{title}{{Lattice QCD at the physical point: Simulation and
  analysis details}},
\newblock \bibinfo{journal}{JHEP} \bibinfo{volume}{08}
  (\bibinfo{year}{2011}{\natexlab{b}}) \bibinfo{pages}{148}.
\bibitem[{Hasenbusch(2001)}]{Hasenbusch:2001ne}
\bibinfo{author}{M.~Hasenbusch},
\newblock \bibinfo{title}{{Speeding up the Hybrid-Monte-Carlo algorithm for
  dynamical fermions}},
\newblock \bibinfo{journal}{Phys. Lett.} \bibinfo{volume}{B519}
  (\bibinfo{year}{2001}) \bibinfo{pages}{177--182}.
\bibitem[{Takaishi and de~Forcrand(2006)}]{Takaishi:2005tz}
\bibinfo{author}{T.~Takaishi}, \bibinfo{author}{P.~de~Forcrand},
\newblock \bibinfo{title}{{Testing and tuning new symplectic integrators for
  hybrid Monte Carlo algorithm in lattice QCD}},
\newblock \bibinfo{journal}{Phys. Rev.} \bibinfo{volume}{E73}
  (\bibinfo{year}{2006}) \bibinfo{pages}{036706}.
\bibitem[{Sexton and Weingarten(1992)}]{Sexton:1992nu}
\bibinfo{author}{J.~C. Sexton}, \bibinfo{author}{D.~H. Weingarten},
\newblock \bibinfo{title}{{Hamiltonian evolution for the hybrid Monte Carlo
  algorithm}},
\newblock \bibinfo{journal}{Nucl. Phys.} \bibinfo{volume}{B380}
  (\bibinfo{year}{1992}) \bibinfo{pages}{665--678}.
\bibitem[{Borsanyi et~al.(2012)}]{Borsanyi:2012zs}
\bibinfo{author}{S.~Borsanyi}, et~al.,
\newblock \bibinfo{title}{{High-precision scale setting in lattice QCD}}
  (\bibinfo{year}{2012}).
\bibitem[{Luscher(2010)}]{Luscher:2010iy}
\bibinfo{author}{M.~Luscher},
\newblock \bibinfo{title}{{Properties and uses of the Wilson flow in lattice
  QCD}},
\newblock \bibinfo{journal}{JHEP} \bibinfo{volume}{1008} (\bibinfo{year}{2010})
  \bibinfo{pages}{071}.
\bibitem[{Endrodi et~al.(2011)Endrodi, Fodor, Katz, and Szabo}]{Endrodi:2011gv}
\bibinfo{author}{G.~Endrodi}, \bibinfo{author}{Z.~Fodor},
  \bibinfo{author}{S.~Katz}, \bibinfo{author}{K.~Szabo},
\newblock \bibinfo{title}{{The QCD phase diagram at nonzero quark density}},
\newblock \bibinfo{journal}{JHEP} \bibinfo{volume}{1104} (\bibinfo{year}{2011})
  \bibinfo{pages}{001}.
\bibitem[{Aoki et~al.(2006{\natexlab{a}})Aoki, Endrodi, Fodor, Katz, and
  Szabo}]{Aoki:2006we}
\bibinfo{author}{Y.~Aoki}, \bibinfo{author}{G.~Endrodi},
  \bibinfo{author}{Z.~Fodor}, \bibinfo{author}{S.~D. Katz},
  \bibinfo{author}{K.~K. Szabo},
\newblock \bibinfo{title}{{The order of the quantum chromodynamics transition
  predicted by the standard model of particle physics}},
\newblock \bibinfo{journal}{Nature} \bibinfo{volume}{443}
  (\bibinfo{year}{2006}{\natexlab{a}}) \bibinfo{pages}{675--678}.
\bibitem[{Aoki et~al.(2006{\natexlab{b}})Aoki, Fodor, Katz, and
  Szabo}]{Aoki:2006br}
\bibinfo{author}{Y.~Aoki}, \bibinfo{author}{Z.~Fodor}, \bibinfo{author}{S.~D.
  Katz}, \bibinfo{author}{K.~K. Szabo},
\newblock \bibinfo{title}{{The QCD transition temperature: Results with
  physical masses in the continuum limit}},
\newblock \bibinfo{journal}{Phys. Lett.} \bibinfo{volume}{B643}
  (\bibinfo{year}{2006}{\natexlab{b}}) \bibinfo{pages}{46--54}.
\bibitem[{Fodor and Katz(2002)}]{Fodor:2001au}
\bibinfo{author}{Z.~Fodor}, \bibinfo{author}{S.~Katz},
\newblock \bibinfo{title}{{A New method to study lattice QCD at finite
  temperature and chemical potential}},
\newblock \bibinfo{journal}{Phys.Lett.} \bibinfo{volume}{B534}
  (\bibinfo{year}{2002}) \bibinfo{pages}{87--92}.
\bibitem[{Fodor and Katz(2004)}]{Fodor:2004nz}
\bibinfo{author}{Z.~Fodor}, \bibinfo{author}{S.~Katz},
\newblock \bibinfo{title}{{Critical point of QCD at finite T and mu, lattice
  results for physical quark masses}},
\newblock \bibinfo{journal}{JHEP} \bibinfo{volume}{0404} (\bibinfo{year}{2004})
  \bibinfo{pages}{050}.
\bibitem[{Bloch and Wettig(2006)}]{Bloch:2006cd}
\bibinfo{author}{J.~C.~R. Bloch}, \bibinfo{author}{T.~Wettig},
\newblock \bibinfo{title}{{Overlap Dirac operator at nonzero chemical potential
  and random matrix theory}},
\newblock \bibinfo{journal}{Phys. Rev. Lett.} \bibinfo{volume}{97}
  (\bibinfo{year}{2006}) \bibinfo{pages}{012003}.
\bibitem[{Gavai and Sharma(2010)}]{Gavai:2009vb}
\bibinfo{author}{R.~Gavai}, \bibinfo{author}{S.~Sharma},
\newblock \bibinfo{title}{{Anomalies at finite density and chiral fermions}},
\newblock \bibinfo{journal}{Phys.Rev.} \bibinfo{volume}{D81}
  (\bibinfo{year}{2010}) \bibinfo{pages}{034501}.
\bibitem[{Narayanan and Sharma(2011)}]{Narayanan:2011ff}
\bibinfo{author}{R.~Narayanan}, \bibinfo{author}{S.~Sharma},
\newblock \bibinfo{title}{{Introduction of the chemical potential in the
  overlap formalism}},
\newblock \bibinfo{journal}{JHEP} \bibinfo{volume}{1110} (\bibinfo{year}{2011})
  \bibinfo{pages}{151}.
\bibitem[{Egri et~al.(2007)Egri, Fodor, Hoelbling, Katz, Nogradi
  et~al.}]{Egri:2006zm}
\bibinfo{author}{G.~I. Egri}, \bibinfo{author}{Z.~Fodor},
  \bibinfo{author}{C.~Hoelbling}, \bibinfo{author}{S.~D. Katz},
  \bibinfo{author}{D.~Nogradi}, et~al.,
\newblock \bibinfo{title}{{Lattice QCD as a video game}},
\newblock \bibinfo{journal}{Comput.Phys.Commun.} \bibinfo{volume}{177}
  (\bibinfo{year}{2007}) \bibinfo{pages}{631--639}.

\end{thebibliography}

\end{document}